\documentclass[prd,twocolumn,nofootinbib,superscriptaddress,amssymb,amsfonts,amsmath,preprintnumbers]{revtex4-2}
\usepackage{graphicx,slashed,xcolor,multirow,bbold,mathtools,sidecap,tikz,bm,ulem,enumitem,booktabs,array}
\usepackage[colorlinks=true,linktoc=page,linkcolor=purple,citecolor=teal,urlcolor=magenta]{hyperref}
\usepackage[compat=1.1.0]{tikz-feynman}
\usepackage{siunitx,adjustbox}

\definecolor{lime}{HTML}{A6CE39}
\DeclareRobustCommand{\orcidicon}{\hspace{-2.1mm}
\begin{tikzpicture}
\draw[lime,fill=lime] (0,0.0) circle [radius=0.13] node[white] {{\fontfamily{qag}\selectfont \tiny ID}}; \draw[white,fill=white] (-0.0525,0.095) circle [radius=0.007]; 
\end{tikzpicture} \hspace{-3.7mm} }
\foreach \x in {A, ..., Z} {\expandafter\xdef\csname orcid\x\endcsname{\noexpand\href{https://orcid.org/\csname orcidauthor\x\endcsname} {\noexpand\orcidicon}}}

\let\emph\textit

\begin{document}

\preprint{PSI-PR-24-06, ZU-TH 09/24, ICPP-79}

\title{
Explaining the $\gamma\gamma+X$ Excesses at $\approx$151.5\,GeV\\ via the Drell-Yan Production of a Higgs Triplet
}

\author{Saiyad Ashanujjaman\orcidA{}}
\email{s.ashanujjaman@gmail.com}
\affiliation{Institute of High Energy Physics, Chinese Academy of Sciences, Beijing 100049, China}
\affiliation{Kaiping Neutrino Research Center, Jiangmen 529399, China}

\author{Sumit Banik\orcidB{}}
\email{sumit.banik@psi.ch}
\affiliation{Physik-Institut, Universität Zürich, Winterthurerstrasse 190, CH–8057 Zürich, Switzerland}
\affiliation{Paul Scherrer Institut, CH–5232 Villigen PSI, Switzerland}

\author{Guglielmo Coloretti\orcidC{}}
\email{guglielmo.coloretti@physik.uzh.ch}
\affiliation{Physik-Institut, Universität Zürich, Winterthurerstrasse 190, CH–8057 Zürich, Switzerland}
\affiliation{Paul Scherrer Institut, CH–5232 Villigen PSI, Switzerland}

\author{Andreas Crivellin\orcidD{}}
\email{andreas.crivellin@cern.ch}
\affiliation{Physik-Institut, Universität Zürich, Winterthurerstrasse 190, CH–8057 Zürich, Switzerland}
\affiliation{Paul Scherrer Institut, CH–5232 Villigen PSI, Switzerland}

\author{Siddharth P.~Maharathy\orcidE{}}
\email{siddharth.m@iopb.res.in}
\affiliation{Institute of Physics, Bhubaneswar, Sachivalaya Marg, Sainik School, Bhubaneswar 751005, India}
\affiliation{Homi Bhabha National Institute, Training School Complex, Anushakti Nagar, Mumbai 400094, India}

\author{Bruce Mellado}
\email{bmellado@mail.cern.ch}
\affiliation{School of Physics and Institute for Collider Particle Physics, University of the Witwatersrand, Johannesburg, Wits 2050, South Africa}
\affiliation{iThemba LABS, National Research Foundation, PO Box 722, Somerset West 7129, South Africa}

\begin{abstract}
The multi-lepton anomalies and searches for the associated production of a narrow resonance indicate the existence of a $\approx$151\,GeV Higgs with a significance of $>5\sigma$ and $>3.9\sigma$, respectively. On the one hand, these anomalies require a sizable branching fraction of the new scalar to $WW$, while on the other hand, no $ZZ$ signal at this mass has been observed. This suggests that the new boson is the neutral component of an $SU(2)_L$ triplet with zero hypercharge. This field leads to a positive definite shift in the $W$ mass, as preferred by the current global fit, and is produced via the Drell-Yan process $pp\to W^*\to \Delta^0\Delta^\pm$. We use the side-bands of the ATLAS analysis~\cite{ATLAS:2023omk} of the associated production of the Standard Model Higgs in the di-photon channel to search for this production mode of the triplet. Since the dominant decays of $\Delta^\pm$ depend only on its mass, the effect in the 22 signal categories considered by ATLAS is completely correlated. We find that the ones most sensitive to the Drell-Yan production of the triplet Higgs show consistent excesses at a mass of $\approx$151.5\,GeV. Combining these channels in a likelihood ratio test, a non-zero Br$[\Delta^0\to\gamma\gamma] = 0.66\%$ is preferred by $\approx$3$\sigma$, supporting our conjecture. 
\end{abstract}
\maketitle

\section{Introduction} 
The Standard Model (SM) of particle physics is the established theoretical description of the fundamental constituents of matter and their interactions. It has been extensively and successfully tested at the precision frontier~\cite{ALEPH:2005ab,HeavyFlavorAveragingGroup:2022wzx,ParticleDataGroup:2022pth} and its last missing particle, the Brout-Englert-Higgs boson~\cite{Higgs:1964ia,Englert:1964et,Higgs:1964pj,Guralnik:1964eu} has been discovered in 2012 at CERN~\cite{Aad:2012tfa,Chatrchyan:2012ufa,CDF:2012laj}. In fact, this $125\,$GeV scalar has properties consistent with the ones predicted by the SM~\cite{Chatrchyan:2012jja,Aad:2013xqa,ATLAS:2016neq,Langford:2021osp,ATLAS:2021vrm}.

However, the minimality of the SM Higgs sector, i.e.~a single $SU(2)_L$ doublet scalar that simultaneously gives masses to the electroweak (EW) gauge bosons and all (fundamental) fermions, is not guaranteed by any theoretical principle or symmetry. Therefore, a plethora of extensions have been proposed in the literature, including the addition of $SU(2)_L$ singlets~\cite{Silveira:1985rk,Pietroni:1992in,McDonald:1993ex}, doublets~\cite{Lee:1973iz,Haber:1984rc,Kim:1986ax,Peccei:1977hh,Turok:1990zg} and triplets~\cite{Konetschny:1977bn,Cheng:1980qt,Lazarides:1980nt,Schechter:1980gr,Magg:1980ut,Mohapatra:1980yp}, as well as countless more models containing multiple new scalars. 

To be viable, such extensions must respect the bounds from electroweak precision observables (particularly the $\rho$ parameter), flavour observables and searches at the Large Hadron Collider (LHC). While flavour observables can be avoided if the new Higgses do not couple directly to SM fermions, the experimental average of the $W$ mass~\cite{ParticleDataGroup:2022pth} even suggests a positive new physics (NP) effect with $3.7\sigma$ significance~\cite{deBlas:2022hdk,Athron:2022isz,Bagnaschi:2022whn}. Furthermore, LHC searches revealed increasing excesses for new scalar bosons at the EW scale~\cite{Crivellin:2023zui}, most pronounced at $\approx$151.5\,GeV~\cite{Crivellin:2021ubm} ($3.9\sigma$). This is consistent with the mass range of 145\,GeV-155\,GeV predicted by a new physics explanation of the multi-lepton anomalies~\cite{vonBuddenbrock:2017gvy}, {\it i.e.}~LHC processes involving multiple leptons and missing transverse momentum in the final state (see Refs.~\cite{Fischer:2021sqw,Crivellin:2023zui} for recent reviews). 

An explanation of the multi-lepton anomalies~\cite{vonBuddenbrock:2015ema,vonBuddenbrock:2016rmr,vonBuddenbrock:2017gvy,vonBuddenbrock:2018xar,Buddenbrock:2019tua,Hernandez:2019geu,vonBuddenbrock:2020ter,Banik:2023vxa}, in particular the very significant disagreement in the top quark differential distributions~\cite{ATLAS:2023gsl,Buddenbrock:2019tua,Banik:2023vxa} ($>5\sigma$), requires the new scalar to decay dominantly to $WW$. In fact, also resonant searches show indications for this decay mode~\cite{CMS:2022uhn,ATLAS:2022ooq,Coloretti:2023wng}, but no $ZZ$ excess at this mass is observed~\cite{CMS:2022dwd,ATLAS:2020rej}. This suggests that the neutral component of the $SU(2)_L$ triplet with hypercharge $Y=0$, which, at tree-level in the absence of mixing, couples to $W$ bosons but not to the $Z$, could be the $\approx$151\,GeV boson. This field, at the same time, predicts a positive definite shift in the $W$ mass (with respect to the SM prediction)~\cite{Chabab:2018ert,FileviezPerez:2022lxp, Cheng:2022hbo,Chen:2022ocr,Rizzo:2022jti,Chao:2022blc,Wang:2022dte,Shimizu:2023rvi,Lazarides:2022spe,Senjanovic:2022zwy,Crivellin:2023gtf,Chen:2023ins,Ashanujjaman:2023etj}. Furthermore, it cannot have tree-level couplings to SM fermions, such that it naturally avoids flavour bounds and has in general suppressed gluon and vector boson fusion production cross-sections, weakening or avoiding LHC bounds.

The triplet Higgs is produced via the Drell-Yan process $pp\to W^*\to \Delta^0\Delta^\pm$ at the LHC, i.e.~together with additional particles. This has the potential to explain the hints for the $\approx$151\,GeV boson, which are most pronounced in associated production channels. Note that these exclusive searches, in which the presence of additional particles or signatures in the final state, together with the ones from which the invariant mass is reconstructed, are required, can significantly improve the new physics sensitivity by reducing the SM background. 

The most extensive and complete search for the associated production of the SM Higgs was performed in Ref.~\cite{ATLAS:2023omk}. In addition to the photon pair, from which the mass of the Higgs is reconstructed, ATLAS searched in its model-independent analysis 22 different signatures, including leptons, jets, missing energy, {\it etc}. While no significant excess for $m_{\gamma\gamma}=125$\,GeV was found (in a single channel), the side-bands of this analysis can be used to search for the associated production of new Higgses in the 110\,GeV--160\,GeV range. 

Therefore, we consider the process $pp\to W^*\to (\Delta^0\to \gamma\gamma)(\Delta^\pm\to XY)$, see Fig.~\ref{fig:ppToHpmH0}. For our mass range, $XY=,tb,WZ,\tau\nu$ (including that $t$, $W$ or $Z$ can be off-shell) are the dominant decay rates. Because $\Delta^\pm$ and $\Delta^0$ turn out to be quasi-degenerate in mass and the branching ratios of $\Delta^\pm$ solely depend on it, Br$(\Delta^0\to \gamma\gamma)$ is the only relevant additional free parameter (for any given mass), making our setup very predictive. We can thus perform a global fit taking the full set of channels of Ref.~\cite{ATLAS:2023omk} to test our conjecture that the neutral component of the $SU(2)_L$ triplet is the (hypothetical) $\approx$151\,GeV boson.

\section{Model and Analysis}

\begin{figure}[t!]
    \centering
    \begin{tikzpicture}[baseline=(current bounding box.center)]
        \begin{feynman}
            \vertex (a);
            \vertex [above left=1.5cm of a] (c) {$p$};
            \vertex [below left=1.5cm of a] (d) {$p$};
            \vertex [right=1.5cm of a] (b) ;
            \vertex [above right=1.5cm of b] (e);
            \vertex [below right=1.5cm of b] (f);
            \vertex [above right=0.75cm of e] (i) {$\gamma$};
            \vertex [below right=0.75cm of e] (j) {$\gamma$};
            \vertex [above right=0.75cm of f] (k) {$X$};
            \vertex [below right=0.75cm of f] (l) {$Y$};
            \diagram{
                (d) -- [fermion] (a) -- [fermion] (c);
                (a) -- [boson, edge label=$W^*$] (b);
                (f) -- [scalar, edge label=$\Delta^\pm$] (b) -- [scalar, edge label=$\Delta^0$] (e);
                (j) -- [boson] (e) -- [boson] (i);
                (l) -- (f) --  (k);
            };
        \end{feynman}
    \end{tikzpicture}
    \caption{Feynman diagram showing the Drell-Yan production of the triplet, $pp\to W^*\to (\Delta^{\pm}\to XY)(\Delta^{0}\to\gamma\gamma)$, which can be searched for in the side-bands of the ATLAS analysis of the associated production of the SM Higgs~\cite{ATLAS:2023omk}. Within our mass range, $XY$ are dominantly $WZ,tb,cs,\tau\nu$ (see Fig.~\ref{fig:xsec}).}
    \label{fig:ppToHpmH0}
\end{figure}
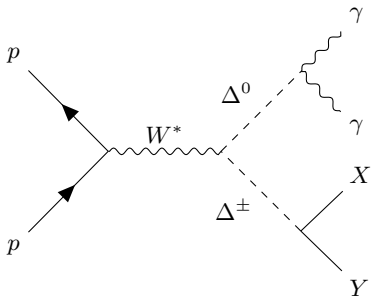

The SM extended with an $SU(2)_L$ triplet scalar with hypercharge $Y=0$ is commonly referred to as the $\Delta$SM~\cite{Ross:1975fq,Gunion:1989ci,Chankowski:2006hs,Blank:1997qa,Forshaw:2003kh,Chen:2006pb,Chivukula:2007koj,Bandyopadhyay:2020otm}. It contains, in addition to the SM Higgs $h$, a charged scalar ($\Delta^\pm$) and a neutral one ($\Delta^0$). In the process of spontaneous electroweak symmetry breaking, it acquires a vacuum expectation value (VEV) $v_\Delta$, which maximally breaks custodial symmetry because it only contributes to the $W$ mass but not to the $Z$ mass.\footnote{For details on the model and its free parameters, see e.g.~\cite{Ashanujjaman:2023etj}.} Therefore, it leads to a positive definite shift in $m_W$ (at tree-level) w.r.t.~the SM prediction~\cite{deBlas:2022hdk,Sirlin:1983ys,Djouadi:1987gn,Avdeev:1994db,Chetyrkin:1995js,Chetyrkin:1995ix,Awramik:2003rn,Degrassi:2014sxa,ParticleDataGroup:2022pth}. This is in agreement with the current global average~\cite{ParticleDataGroup:2022pth}, which indicates a positive effect of $\approx$20\,MeV with a significance of $3.7\sigma$~\cite{deBlas:2022hdk}.\footnote{The global fit is based on the LHC~\cite{ATLAS:2023fsi,LHCb:2021bjt}, Tevatron~\cite{CDF:2022hxs} and LEP~\cite{ALEPH:2013dgf} results. While there is a tension between the most precise measurement from CDF-II with the LHC combination, which leads to an enlargement of the error, notice that, also excluding the CDF-II result, a positive shift is preferred at $\approx$2$\sigma$.} While the exact value is immaterial for this work, the $W$ mass implies $v_\Delta=\mathcal{O}({\rm GeV})$.

Since the mass splitting between $\Delta^\pm$ and $\Delta^0$ is of the order of $v_\Delta$ (for the scalar self-couplings of $\mathcal{O}(1)$), we can assume them to be degenerate to a good approximation, {\it i.e.}~$m_{\Delta^\pm}\approx m_{\Delta^0}\approx m_{\Delta}$. The mixing angle $\alpha$ between $h$ and $\Delta^0$ impacts the branching ratios of $\Delta^0$. However, we are only interested in $\Delta^0\to\gamma\gamma$ which is loop induced and depends not only on $\alpha$ but also critically on $m_{\Delta^0}-m_{\Delta^\pm}$ (because it is related to the $\Delta^0\Delta^\pm\Delta^\mp$ coupling). Therefore, we can subsume these parameters within Br$(\Delta^0\to\gamma\gamma)$ such that it is the only free parameter in our analysis (for any given mass). 

\begin{table}[t!]
\begin{center}
\begin{tabular}{m{1.5cm}m{2.5cm}m{4.2cm}}
\hline
\\[-.1cm]
Target & Region & Detector level \\
\\[-.1cm]
\hline
\\[-.1cm]
\multirow{2}{1.5cm}{High jet activity} & $\ge 4j$ & $n_{jet}~\ge 4$ , $|\eta_{jet}|~<2.5$  \\
&&\\
\\[-.1cm]
\hline
\\[-.1cm]
\multirow{3}{1.5cm}{Top} & $\ell b$ & $n_{\ell=e,\mu} \ge 1$, $n_{b\text{-jet}} \ge 1$ \\
& $t_\text{lep}$ &  $n_{\ell=e,\mu}=1$, $n_\text{jet}=n_{b\text{-jet}}=1$ \\
& $t_\text{had}$ & $n_{\ell=e,\mu}=0$, $n_\text{jet}=3$,$n_{b\text{-jet}}= 1$ \\
\\[-.1cm]
\hline
\\[-.1cm]
\multirow{2}{1.5cm}{Lepton} & $\geq$1$\ell$ & $n_{\ell=e,\mu}\ge1$ \\
& $2\ell$ & $ee,\mu\mu$ or $e\mu$ \\
& $\geq$3$\ell$ & $n_{\ell=e,\mu}\ge3$ \\
\\[-.1cm]
\hline
\\[-.1cm]
\multirow{3}{4cm}{$E_\text{T}^\text{miss}$} & $E_\text{T}^\text{miss} > 100 \text{ GeV}$  & $E_\text{T}^\text{miss} > 100 \text{ GeV}$ \\
& $E_\text{T}^\text{miss} > 200 \text{ GeV}$  & $E_\text{T}^\text{miss} > 200 \text{ GeV}$  \\
& $E_\text{T}^\text{miss} > 300 \text{ GeV}$  & $E_\text{T}^\text{miss} > 300 \text{ GeV}$ \\
\\[-.1cm]
\hline
\end{tabular}
\end{center}
\caption{The signal regions of the ATLAS analysis~\cite{ATLAS:2023omk} that are most relevant for the Drell-Yan production of the scalar triplet.}
\label{categories}
\end{table}

\begin{figure*}[t!]
\centering
\includegraphics[scale=0.85]{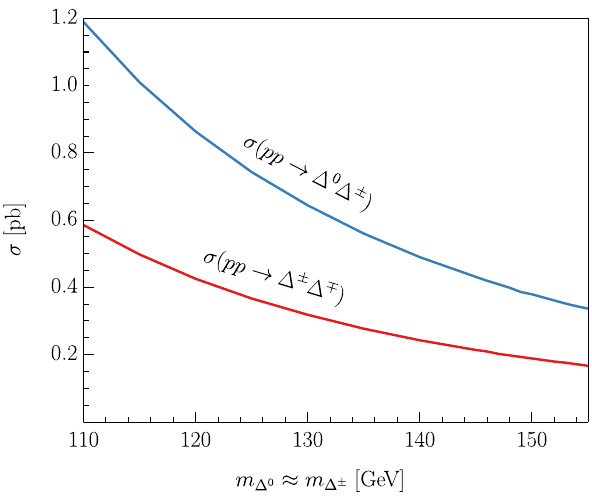}
\hspace{0.3cm}
\includegraphics[scale=0.85]{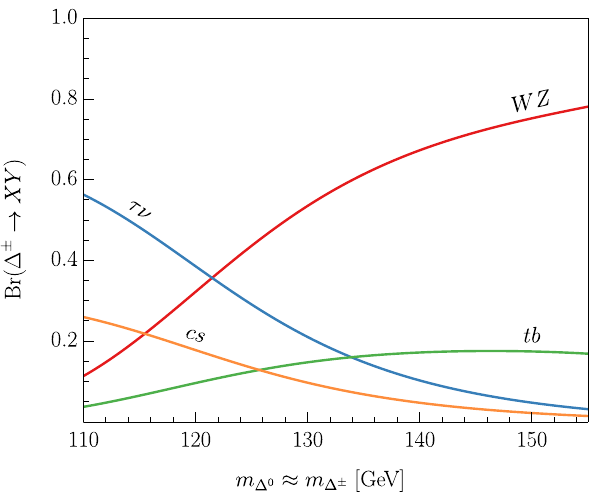}
\caption{Left: Production cross-section for $pp\to \Delta^0 \Delta^\pm$ and $pp\to \Delta^\pm \Delta^\mp$ as a function of the triplet mass including the NNLL and NLO QCD correction factor of Refs.~\cite{Ruiz:2015zca,AH:2023hft}. Right: Dominant branching ratios of the charged component $\Delta^\pm$ as a function of the mass.}
\label{fig:xsec}
\end{figure*}

The production cross-sections for the Drell-Yan processes, $pp\to W^*\to \Delta^0\Delta^\pm$ and $pp\to Z^*,\gamma^*\to \Delta^\mp\Delta^\pm$ (where the asterisk denotes that the particle is off-shell), as a function of $m_{\Delta}$ are given in Fig.~\ref{fig:xsec} (left). Here we include the next-to-next-to leading log (NNLL) and NLO QCD correction factor of Refs.~\cite{Ruiz:2015zca,AH:2023hft} (see also Ref.~\cite{Djouadi:1999ht} for QCD corrections to charged Higgs production). Note that this is the dominant production mode in the limit of $\alpha\approx 0$ and $v_\Delta\ll v$ (where $v$ is the SM VEV). These processes were studied in detail in Ref.~\cite{Butterworth:2023rnw} finding that 4-lepton searches can only exclude masses between $\approx160\,$GeV and $\approx200\,$GeV.\footnote{In fact, Ref.~\cite{Butterworth:2023rnw} used $m_{\Delta^0}=m_{\Delta^\pm}=150\,$GeV as a benchmark point.} However, the ATLAS analysis of the associated production of the SM Higgs~\cite{ATLAS:2023omk} sensitive to $pp\to W^*\to \Delta^0\Delta^\pm$ with $\Delta^0\to\gamma\gamma$ shown in Fig.~\ref{fig:ppToHpmH0} was not included. 

To determine the efficiency (including cuts and acceptance, {\it etc.}) we simulated $pp\to W^*\to (\Delta^\pm\to XY)(\Delta^0\to \gamma\gamma)$ using {\tt MadGraph5aMC@NLO} with the parton shower performed by {\tt Pythia8.3}~\cite{Sjostrand:2014zea} and carried out the detector simulation for the ATLAS detector~\cite{ATLAS:2023omk} with {\tt Delphes}~\cite{deFavereau:2013fsa}. The UFO model file at NLO for the $\Delta$SM was built using {\tt FeynRules}~\cite{Degrande:2011ua,Alloul:2013bka, Degrande:2014vpa}.\footnote{To reduce the computational resources needed for our scan, we performed the simulations at leading order but rescaled the production cross-section to account for the NLO and NNLL effect. Furthermore, we simulated the 4-jet category at NLO, where gluon radiation is particularly important, and included it via a correction factor of 1.2 in the extraction of the efficiency.} Note that the dominant branching fractions of $\Delta^\pm$ are induced at tree-level by $v_\Delta$ such that the dependence on it cancels. Therefore, we can show them as a function of $m_{\Delta}$ in Fig.~\ref{fig:xsec} (right). The dominant final states for the mass range of interest are $XY=\tau\nu,cs,tb$ and $WZ$ (including implicitly that $t$, $W$ and $Z$ can be off-shell). 

Taking into account that $Z$ and $W$ bosons decay to lepton, missing energy and jets, we expect that the ATLAS signal regions targeting leptons, missing transverse energy ($E_T^{\rm miss}$) and high jet activity are the most important ones. Furthermore, at higher values of $m_{\Delta}$, the categories addressing top quarks become relevant.  

To obtain bounds on Br$(\Delta^0\to \gamma\gamma)$ we perform a likelihood-ratio test using Poisson statistics. This means that we take the theory prediction for the number of events in a bin $i$ as the mean of the Poisson distribution and calculate the ratio
\begin{equation}
    \mathcal{L}_R=\Pi_i\left[\mathcal{L}(N_i^{\rm SM},N_i^{\rm exp})/\mathcal{L}(N_i^{\rm NP},N_i^{\rm exp})\right],
\end{equation}
where $\mathcal{L}$ is the likelihood described by the Poisson distribution, $N_i^{\rm SM,NP}$ is the number of expected events (which is a real number) in the SM (including the SM Higgs) or our $\Delta$SM and $N_i^{\rm exp}$ is the number of events measured by ATLAS. Because $\mathcal{L}(N_i^{\rm NP},N_i^{\rm exp})$ is a function of $m_{\Delta}$ and Br$(\Delta^0\to \gamma\gamma)$, we can maximize it w.r.t.~the latter for any given mass. This then corresponds to the maximum likelihood $\mathcal{L}_R^{\rm max}$ and thus to the best-fit value for Br$(\Delta^0\to \gamma\gamma)$. Note that we allow for an unphysical negative branching ratio to take into account the effect of downward fluctuations of the background. The 68\% and 95\% confidence level (CL) regions are calculated in the asymptotic limit (where $\chi^2=-2 {\rm ln}(\mathcal{L}_R^{\rm max})$) by requiring that $\chi^2=1$ and $\chi^2=4$, respectively.

\begin{figure*}[htb!]
\centering
\includegraphics[scale=0.85]{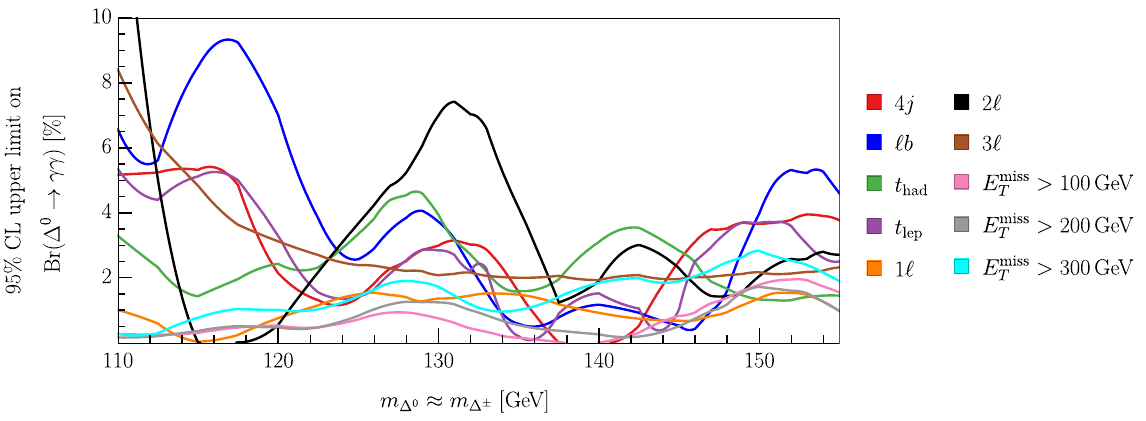}
\caption{95\% CL upper limits on Br($\Delta^0 \to \gamma\gamma$) for the 10 most relevant categories. Note that except for the $t_{had}$ and $3\ell$ signal regions, all categories show a weakening of the limit around 151\,GeV.}
\label{fig:exclusion_single}
\end{figure*}

\section{Results}
\begin{figure}[t!]
\includegraphics[scale=0.85]{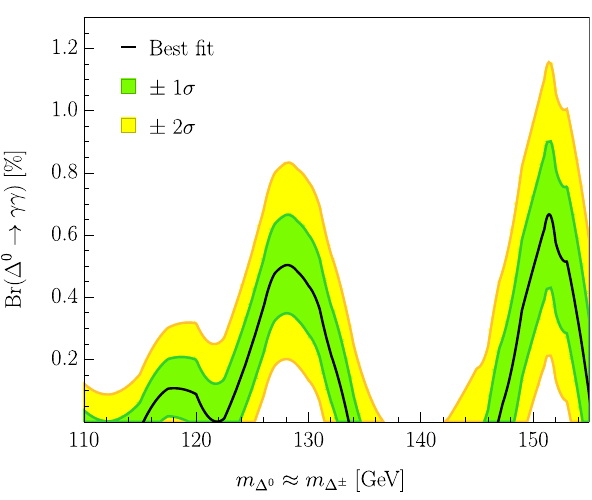}
\caption{Statistical combination of the relevant channels. Note that a non-zero branching ratio of $\Delta^0\to \gamma\gamma$ is preferred at both $\approx$127\,GeV ($3.6\sigma$) and $\approx$151\,GeV ($3\sigma$).}
\label{fig:exclusion_comb}
\end{figure}

We find that among the 22 categories provided by ATLAS~\cite{ATLAS:2023omk}, the ten detained in Table~\ref{categories} give the most relevant constraints on Br$(\Delta^0\to \gamma\gamma)$. The corresponding 95\% CL bounds on Br$(\Delta^0\to \gamma\gamma)$ are shown in Fig.~\ref{fig:exclusion_single}. Interestingly, all relevant distributions, except $t_{\rm had}$, show a weakening of the limit around a mass of $\approx$150\,GeV. Furthermore, a similar weakening of the limits is observed between 125\,GeV and 130\,GeV. Combing all ten signal regions in Fig.~\ref{fig:exclusion_single} by taking the product of the likelihood ratios, we obtain the best-fit value for Br$(\Delta^0\to \gamma\gamma)$ as a function of the mass, as well the 68\% CL and 95\% CL regions, as shown in Fig.~\ref{fig:exclusion_comb}.\footnote{Note that there is a moderate correlation between the signal regions, in particular between the different $E_T^{\rm miss}$ categories. However, we checked that removing two of the three $E_T^{\rm miss}$ signal regions from the fit has a very small impact on the final result.} One can see that there is a preference for a non-zero signal both at around 125--130\,GeV and at $\approx$151.5\,GeV. The latter has a significance of $3\sigma$ and shows the consistency of data with the assumption that the (hypothetical) $\approx151.5$\,GeV Higgs is the neutral component of a triplet with $Y=0$. 

Furthermore, the excess in the 125\,GeV--130\,GeV range, with a significance of $3.5 \sigma$ at $\approx$ 128\,GeV, could point towards the enhanced associated production of the SM Higgs, since at 125\,GeV the significance is still $1.8 \sigma$. Finally, note that, in principle, the background would be lower if it were refitted in the presence of one or more additional resonances. This would, particularly for our two-peak structures with a deficit in events between them, further increase the significance.

\section{Conclusions and Outlook}

The multi-lepton anomalies ~\cite{vonBuddenbrock:2015ema,vonBuddenbrock:2016rmr,vonBuddenbrock:2017gvy,Buddenbrock:2019tua,Coloretti:2023wng}, in particular the differential $t\bar t$ distributions~\cite{Coloretti:2023wng} ($> 5 \sigma$), suggest the existence of a new scalar with a mass in the 145--155\,GeV with a sizable branching fraction to $WW$. This mass range is in agreement with the indications for a narrow resonance in $\gamma\gamma$ and $Z\gamma$ final states at $\approx$151\,GeV ($>3.9 \sigma$~\cite{Crivellin:2021ubm,Bhattacharya:2023lmu}). However, since no excess in $ZZ\to 4\ell$ is observed~\cite{CMS:2022dwd,ATLAS:2020rej}. This motivates the hypothesis that the $\approx$151\,GeV Higgs could be the neutral component of an $SU(2)_L$ triplet with $Y=0$, because it only couples to $W$ bosons at tree-level (in the absence of mixing) but not to the $Z$. Furthermore, this field predicts a positive shift in $m_W$, as preferred by the current global fit~\cite{deBlas:2022hdk,Bagnaschi:2022whn}. 

For small mixing angles, the main production mechanism of $\Delta^0$ at the LHC is associated production together with the charged component $\Delta^\pm$  via the Drell-Yan process $pp\to W^*\to \Delta^0\Delta^\pm$. In this article, we use the sidebands of the model-independent ATLAS analysis~\cite{ATLAS:2023omk} of the associated production of the SM Higgs in the di-photon channel to search for this process. Because the leading branching fractions of $\Delta^\pm$ only depend on its mass, and it must be quasi-degenerate with the neutral component ($m_{\Delta^\pm}\approx m_{\Delta^0}\approx m_{\Delta}$), Br$(\Delta^0\to\gamma\gamma)$ is the only free parameter of interest (for a given mass). Within the mass range 110-156\,GeV, ten out of the 22 signal regions of the ATLAS analysis~\cite{ATLAS:2023omk} give relevant constraints on Br$(\Delta^0\to\gamma\gamma)$. Furthermore,  9 of them show a weakening of the 95\% CL limit at $\approx$150\,GeV. In fact, combining all channels in a likelihood ratio test leads to the preference of non-zero Br$(\Delta^0\to\gamma\gamma)$, with a significance of $3\sigma$ for $m_{\Delta}$$\approx$151.5\,GeV and a best-fit value of Br$(\Delta^0\to\gamma\gamma)=0.66\%$. This shows that the indications for the associated production of the $\approx$151\,GeV scalar are consistent with the assumption that it is the neutral component of the triplet.

Moreover, a similar excess is obtained near the SM Higgs boson mass. While it is most pronounced at $128\,$GeV ($3.5 \sigma$), it could also originate from the associated production of the SM Higgs. In this case, the significance reduces to $1.8 \sigma$. This option is particularly interesting since it was already pointed out in Ref.~\cite{Hernandez:2019geu} that there are hints for a non-zero new physics contribution to $h+W$, which constitutes an overlapping signal. Furthermore, the enhanced associated production of the SM Higgs is expected in the $\Delta$2HDMS~\cite{Coloretti:2023yyq}, which was recently proposed as a combined explanation of the multi-lepton anomalies, the hints for new Higgses at 152\,GeV and 95\,GeV and the resonant $t\bar t$ excess at 400\,GeV~\cite{ATLAS:2022rws}. 

While in this article, we accumulated evidence that the (hypothetical) $\approx$151\,GeV Higgs is the neutral component of the triplet with $Y=0$, this does not exclude a more complicated scalar sector, and additional production mechanisms are required to explain the multi-lepton anomalies. To disentangle them, we are eagerly waiting for the analysis of LHC run-3 data, and a comprehensive analysis of associated Higgs production by CMS, in the spirit of Ref.~\cite{ATLAS:2023omk}, would be very welcome. Furthermore, the Drell-Yan production of $\Delta^\pm\Delta^\mp$ can lead to $WWW$-like signatures, where an $\approx$2$\sigma$ higher cross-section than expected in the SM is observed~\cite{CMS:2020hjs,ATLAS:2022xnu}. Finally, we remark that the charged component of the triplet could be examined with great precision at future $e^+e^-$ accelerators, such as the Circular Electron-Positron Collider (CEPC)~\cite{CEPCStudyGroup:2018ghi,An:2018dwb}, the Compact Linear Collider (CLIC)~\cite{CLICdp:2018cto}, the Future Circular Collider (FCC-ee)~\cite{FCC:2018evy,FCC:2018byv} and the International Linear Collider (ILC)~\cite{ILC:2013jhg,Adolphsen:2013jya}.

\begin{acknowledgments}
The work of A.C.~is supported by a professorship grant from the Swiss National Science Foundation (No.\ PP00P21\_76884). B.M.~gratefully acknowledges the South African Department of Science and Innovation through the SA-CERN program, the National Research Foundation, and the Research Office of the University of the Witwatersrand for various forms of support. The work of S.A.~is partially supported by the National Natural Science Foundation of China under grant No.~11835013. S.A. and S.P.M.~acknowledge using the SAMKHYA: High-Performance Computing Facility provided by the Institute of Physics, Bhubaneswar.
\end{acknowledgments}

\bibliographystyle{utphys}

\bibliography{bib}

\end{document}